\documentclass[twocolumn]{pazhb_eng}

\usepackage{graphicx}
\usepackage{amsmath}
\usepackage{rotating}
\usepackage{lscape}
\usepackage{ifpdf}
\usepackage{breqn}
\usepackage[hyphens]{url}
\newcommand {\be}{\begin{equation}}
\newcommand {\ee}{\end{equation}}

\newcommand{\angstrom}{\mbox{\normalfont\AA}}

\def\dgr{\hbox{$^\circ$}}

\begin{document}
\journalinfo{2019}{45}{12}{836}[846]

\title{Optical Identification of Four Hard X-ray Sources from the INTEGRAL
Sky Surveys}
\author{\bf D.I. Karasev$^1$\email{dkarasev@iki.rssi.ru}, S.Yu.Sazonov$^1$, A. Yu. Tkachenko$^1$, G. A. Khorunzhev$^1$, R. A. Krivonos$^1$, P. S. Medvedev$^1$, I.A.Zaznobin$^1$, I. A. Mereminskiy$^1$, R. A. Burenin$^1$, M. N. Pavlinsky$^1$ and M. V. Eselevich$^2$\\
$^1$\it{Space Research Institute, Moscow, Russia\\}
$^2$\it{Institute of Solar–Terrestrial Physics, Russian Academy of Sciences, Siberian Branch, P.O. Box 4026, Irkutsk, 664033 Russia}
}
\shortauthor{}
\shorttitle{}
\submitted{23 October 2019}

\begin{abstract}

We continue the study begun in Karasev et al. (2018) and present the results of our optical
identifications of four hard X-ray sources from the INTEGRAL sky surveys. Having first improved the positions of these objects in the sky with the X-ray telescope (XRT) of the Swift observatory, we have identified their counterparts using optical and infrared sky survey data. Then, we have obtained optical spectra for the supposed counterparts with the RTT-150 Russian-Turkish telescope and the AZT-33IK telescope. This has allowed the nature of the objects under study to be established. The sources IGR\,J11079+7106 and IGR\,J12171+7047 have turned out to be extragalactic in nature and be Seyfert
1 and 2 galaxies, respectively, with the second object being characterized by a large absorption column density. The source IGR\,J18165--3912 is most likely an intermediate polar with a very high luminosity. The fourth source, IGR\,J20596+4303, is a chance superposition of two objects -- a Seyfert 2 galaxy and a cataclysmic variable.

\medskip
\end{abstract}

\section*{INTRODUCTION}

The INTEGRAL observatory (Winkler et al. 2003) has operated in orbit and performed sky observations in the hard X-ray energy range (above 20 keV) for 17 years. In this period a high sensitivity has been achieved in various sky regions, including the Galactic
plane (Revnivtsev et al. 2004, 2006; Molkov et al. 2004; Krivonos et al. 2012, 2017; Bird et al. 2016) and a number of extragalactic fields (Grebenev et al. 2013; Mereminskiy et al. 2016), which has led to the discovery of several hundred new hard X-ray sources
(see, e.g., Krivonos et al. 2007, 2012; Bird et al. 2016). The completeness of the catalog of sources detected by the INTEGRAL observatory is very high owing to the big efforts made to identify them in the soft X-ray, visible, and infrared (IR) bands (see, e.g., Masetti et al. 2007, 2010; Tomsick et al. 2009, 2016; Malizia et al. 2010).
In this paper, using the previously accumulated experience (see, e.g., Bikmaev et al. 2006, 2008; Burenin et al. 2008, 2009; Lutovinov et al. 2010;
Karasev et al. 2018), sky survey data in various wavelength ranges, and additional spectroscopic observations with the RTT-150 and AZT-33IK optical telescopes, we investigated four hard X-ray sources: IGR\,J11079+7106, IGR\,J12171+7047 discovered
during a deep extragalactic survey (Mereminskiy et al. 2016) and IGR\,J18165--3912, IGR\,J20596+4303 first detected in a survey of the Galactic plane and its neighborhoods $|b|<17^\circ$(Krivonos et al. 2017).

\begin{table*}[h]
\centering
\footnotesize{
 \begin{flushleft}  
   \caption{Soft X-ray counterparts of the investigated sources}
 \end{flushleft}  
   
\resizebox{18cm}{!} {   
   \begin{tabular}{c|c|c|c|c|c|c}
     \hline
     \hline
     \rule{0cm}{0.5cm}
          Name &  RA     & Dec     & Localization    & Flux (2-10 keV) & Flux (17-60 keV){**} & Notes\\
              & (J2000) & (J2000) & accuracy  & $\times10^{-12}$,  & $\times10^{-11}$,&\\
              &         &         & (90\%), arcsec      & erg s$^{-1}$ cm$^{-2} $& erg s$^{-1}$ cm$^{-2} $&\\ [0.25cm]

    \hline
\rule{0cm}{0.5cm}
     IGR\,J11079+7106    &  11$^h$07$^m$47$^s$.95 &  $+71$\dgr$05$\arcmin$31$\arcsec$.4$ & 1.6 & ${2.26\pm0.31}$ & $0.72\pm0.10$ &Swift\,J1107.8+7107 \\ [0.25cm]
%       \hline
\rule{0cm}{0.5cm}
     IGR\,J12171+7047   &  12$^h$17$^m$26$^s$.23 &  $+70$\dgr$48$\arcmin$05$\arcsec$.4$ & 4.5 & $ 0.12\pm0.04$ & $1.00\pm0.15$ &NGC\,4250\\ [0.25cm]
     
%     \hline
\rule{0cm}{0.5cm}
    IGR\,J18165--3912$^*$     &  18$^h$16$^m$35$^s$.95 &  $-39$\dgr$12$\arcmin$43$\arcsec$.9$ & 2.9 & ${1.56\pm0.23}$ & $0.61\pm0.09$ &\\ [0.25cm]

%          \hline
\rule{0cm}{0.5cm}
      IGR\,J20596+4303  &   &   &  & & $0.57\pm0.08$&\\ [0.25cm]
      
      &  21$^h$00$^m$00$^s$.96 &  $+43$\dgr$02$\arcmin$09$\arcsec$.6$ & 2.2 & ${1.25\pm0.20}$ & &1SXPS\,J210000.9+430208\\ [0.5cm]
     
     &  20$^h$59$^m$15$^s$.76 &  $+43$\dgr$01$\arcmin$06$\arcsec$.1$ & 2.1 & ${3.13\pm0.17}$ & &1SXPS\,J205915.6+430105 \\ [0.5cm]
      
          \hline
       			
     \hline	

     \hline

    \end{tabular}
    
    }
     
     \begin{flushleft}   
     *  From the XRT/Swift data requested by our group.
    
     **  From the INTEGRAL data (Mereminskiy et al. 2016; Krivonos et al. 2017).
     \end{flushleft}   
    }

\end{table*}

%================================================================
\begin{figure*}[t!]
\centering
\includegraphics[width=0.85\columnwidth,trim={0cm 0 0cm 0cm},clip]{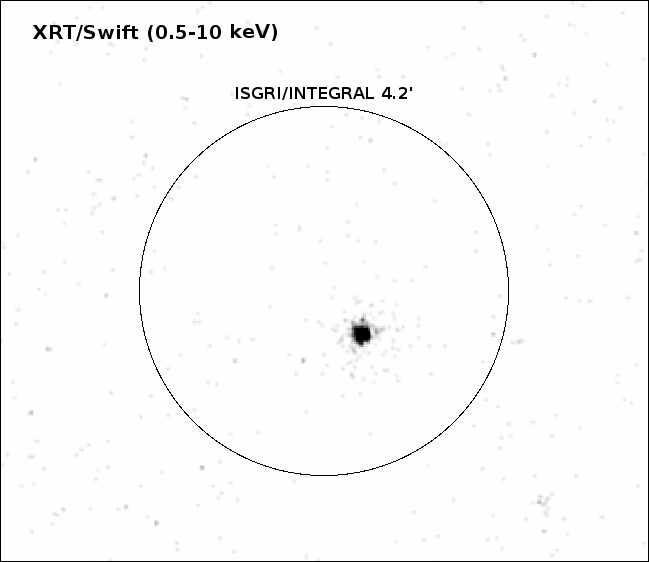}
\includegraphics[width=0.92\columnwidth,trim={-2cm 0cm 0cm 0cm},clip]{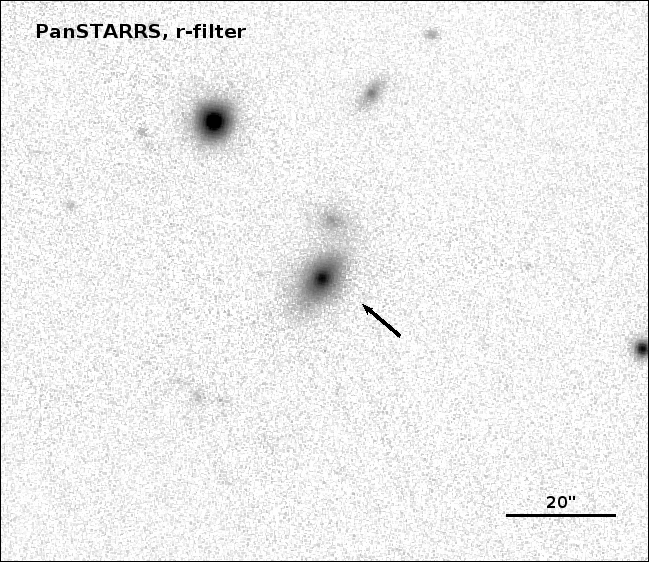}
\includegraphics[width=0.93\columnwidth,trim={0cm 7cm 0cm 3cm},clip]{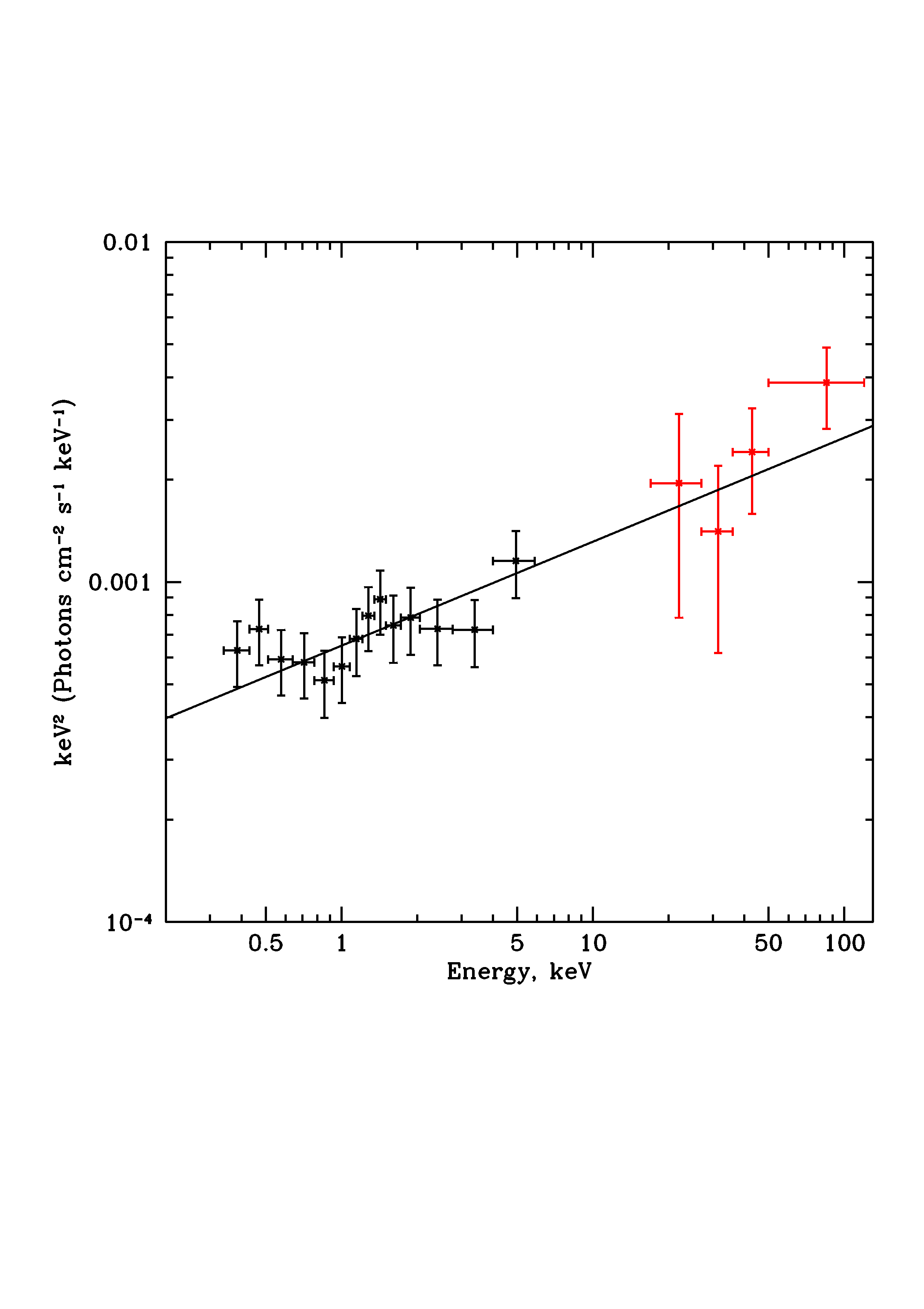}
\includegraphics[width=0.95\columnwidth,trim={0cm 0cm 0cm 3cm},clip]{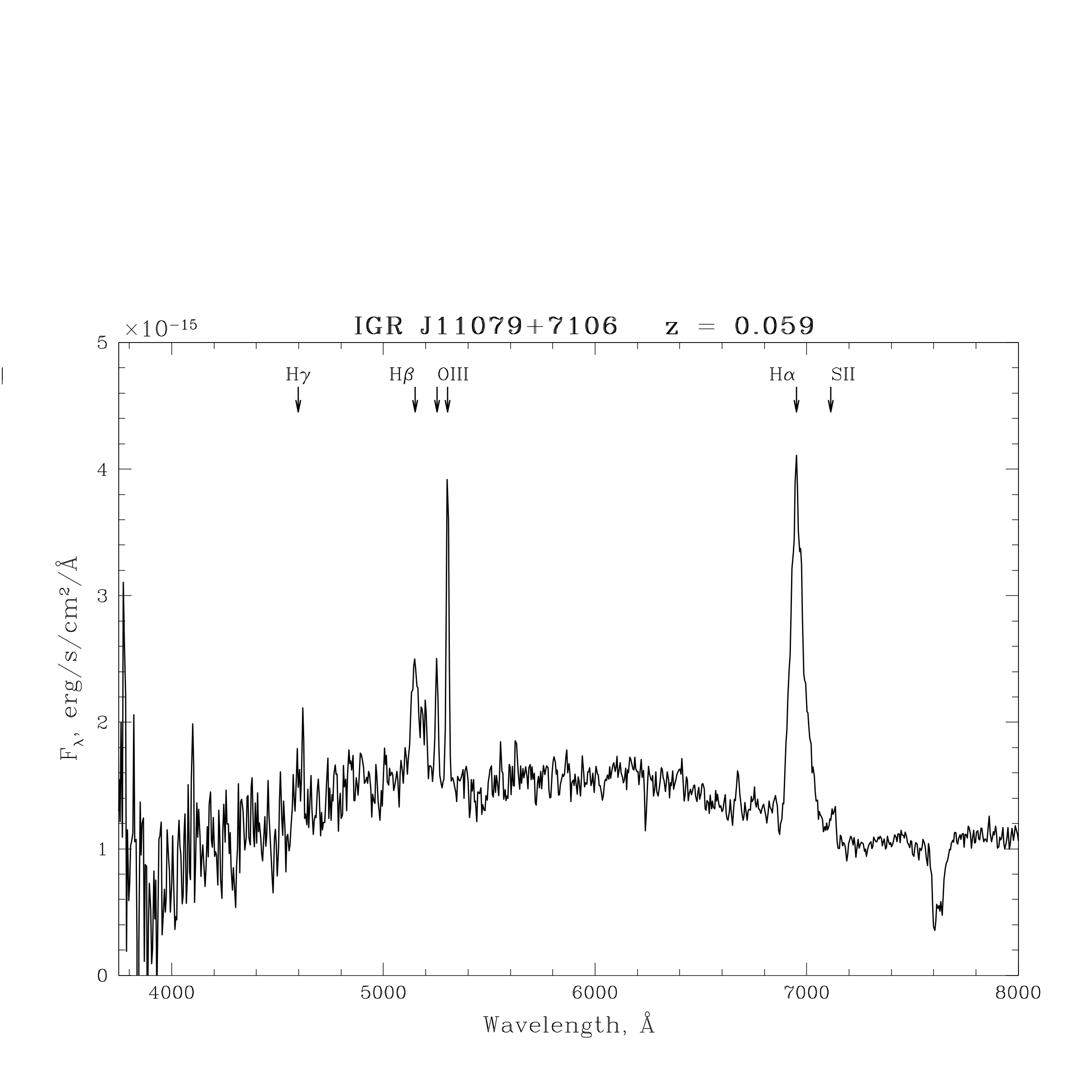}

\caption{Upper row: images of the sky region containing IGR\,J11079+7106 obtained from the XRT/Swift data (left) and in the PanSTARRS sky survey in the r-filter (right). On the left panel the INTEGRAL error circle of the source is indicated by the solid circle. The arrow marks the assumed optical counterpart of the investigated source. Lower row (left): the broadband X-ray spectrum of the source from the XRT/Swift (black dots at energies below 10 keV) and IBIS/INTEGRAL (red dots at energies above 10 keV) data. The solid line indicates a power-law fit to the spectrum. Lower row (right): the RTT-150 optical spectrum of the source. The most significant emission lines are marked.
}  \label{fig:IGR1}
\end{figure*}
%================================================================
 \section*{INSTRUMENTS AND DATA}

Coded-aperture X-ray instruments allow the positions of sources to be determined with an accuracy that, as a rule, is insufficient for their unequivocal optical identification. Therefore, to improve the localization of the objects from our sample, we used publicly
accessible sky observations with the grazing incidence X-ray telescope (XRT) of the Neil Gehrels Swift observatory (hereafter simply the Swift observatory). 

Some of the investigated objects have been detected for the first time and have not been observed previously by other observatories. To improve the localization accuracy and to optically identify these sources, we additionally requested the XRT/Swift
soft X-ray observations. Note that for all objects, based on the significance
of their detection, we took the INTEGRAL position accuracy to be $4.2'$ (at 95.4\% confidence; Krivonos et al. 2007, 2017). This is a fairly large uncertainty
and, therefore, we ran into ambiguity in choosing a soft X-ray counterpart of the hard X-ray source. The spectra of the hard X-ray sources used here were reconstructed
from the INTEGRAL observations over $\sim14$ years (from December 2002 to March 2017)
using the software developed at the Space Research Institute of the Russian Academy of Sciences (Churazov et al. 2005, 2014; Krivonos et al. 2010).

The XRT/Swift data were processed with the corresponding software\footnote{http://swift.gsfc.nasa.gov} of HEASOFT 6.22\footnote{https://heasarc.nasa.gov/lheasoft/}. The positions of the objects in the XRT/Swift images and their localization accuracy were determined with standard recommended procedures and algorithms\footnote{http://www.swift.ac.uk/user\_objects/} (Goad et al. 2007; Evans et al. 2009). 
The X-ray spectra of the sources were fitted using the XSPEC software.

The positions of the sources in the optical and
near-IR bands presented in the paper were taken from
the publicly accessible catalogs of the PanSTARRS\footnote{https://panstarrs.stsci.edu},
VHS ESO\footnote{http://horus.roe.ac.uk/vsa/} and SkyMapper\footnote{http://skymapper.anu.edu.au} sky surveys.  We investigated the mid-IR colors of the objects based on data from the publicly accessible catalogs of the WISE space observatory.

The distances to the optical counterparts of some of the investigated objects were estimated from the data of the Gaia\footnote{https://sci.esa.int/web/gaia} space observatory specially processed by Bailer-Jones et al. (2018) (see the I/347 catalog of the Vizier\footnote{http://vizier.u-strasbg.fr/viz-bin/VizieR}). 

The spectroscopy for a number of objects from the sample was performed at the Russian-Turkish 1.5-m telescope (RTT-150) with the medium- and low resolution TFOSC\footnote{http://hea.iki.rssi.ru/rtt150/ru/index.php?page=tfosc} spectrograph For this purpose, we used the N15 grism, which gives the widest wavelength
range (3500--9000$ \angstrom$) and the greatest quantum efficiency. The spectral resolution was $\approx12 \angstrom$ (FWHM).

Apart from RTT-150, for our spectroscopy we used the 1.6-m AZT-33IK telescope of the Sayan observatory (Kamus et al. 2002) equipped with the ADAM spectrograph with a 2-arcsec slit and a VPHG600R grating. The spectral resolution of the instrument is $4.3 \angstrom$ (FWHM) in the wavelength range 3700--7340$ \angstrom$\,\,(Afanasiev et al. 2016; Burenin et al. 2016).

All of the spectroscopic observations were processed in a standard way using the IRAF\footnote{http://iraf.noao.edu} software and our own additional software package. To calculate the photometric distances $D_{L}$  to the extragalactic sources from their redshifts, we used the $\Lambda$CDM  cosmological model with the following parameters: $H_0$ = 67.8 and $\Omega_M$= 0.308 (Planck Collaboration 2016).

\section*{IDENTIFICATION OF SOURCES}

Basic data on the sources investigated in this paper are presented in Table 1. It provides the names of the sources, the coordinates of their supposed soft X-ray counterparts from the Swift data, the localization accuracy, and the fluxes in the 2--10 keV and 17--60 keV energy bands.

Detailed information about the properties and presumed nature of each of the objects listed in Table 1 is given below.

\bigskip
%\begin{landscape}

%\end{landscape}

%%%%%%%%%%%%%%%%%%%%%%%%%%%%%%%%%%%%%%%%%%%%%%%%%%%%                       1                            %%%%%%%%%%%%%%%%%%%%%%%%%%%%%%%%%%%%%%%%%%%%%%%%%%%%%%%

\subsection*{IGR\,J11079+7106}

According to the Swift archival data, only one soft X-ray source, Swift\,J1107.8+7107 (Evans
et al. 2014), falls into the IBIS/INTEGRAL error circle. It was localized with a high accuracy (see Table 1) from the XRT/Swift data. 

We obtained the XRT X-ray spectrum of Swift\,J1107.8+7107 and combined it with the IBIS/INTEGRAL spectrumof the hard X-ray source IGR\,J11079+7106 (Fig. 1). It turned out that the fluxes from the sources in the soft and hard X-ray bands agree well between themselves, while the broadband spectrum could be fitted ($\chi^2=0.65$ per degree of freedom for 16 degrees of freedom) by a power law with a slope (photon index) of $1.70\pm0.05$ (Fig. 1). Such a spectral model is typical primarily for active galactic nuclei (AGNs).

Using an accurate localization of the assumed soft X-ray counterpart and the PanSTARRS sky survey data, we managed to determine the optical counterpart of the source (Fig. 1). This is an extended object with an apparent magnitude $r\simeq17.0$. We performed its spectroscopy with the RTT-150 telescope of the TUBITAK Observatory. Its spectrum (Fig. 1) clear shows a broad (FWHM = 3272 km s$^{-1}$ corrected for the spectral resolution of the instrument), redshifted H$\alpha$ emission line and a number of other characteristic
emission lines (in particular, [OIII] and [SII]), which unambiguously point to the fact that this object is a Seyfert 1 AGN at redshift $z=0.059\pm0.001$ ($D_{L}\simeq272.6$ Mpc).

Note that Swift\,J1107.8+7107 was previously studied by Stephen et al. (2018). However, the position of the X-ray object was determined with a much lower accuracy ($\sim10''$) and differs from the one presented in this paper by $\approx7.5''$. Nevertheless, the authors managed to correctly establish the optical counterpart and to obtain a rough estimate of its redshift ($\approx0.06$). However, Stephen et al. (2018) did not reach any conclusions about the AGN type. The new information obtained about the optical and X-ray (soft and hard X-ray luminosity) properties of IGR\,J11079+7106 and the remaining objects considered below is given in Table 2.

%%%%%%%%%%%%%%%%%%%%%%%%%%%%%%%%%%%%%%%%%%%%%%%%%%%                      2                          %%%%%%%%%%%%%%%%%%%%%%%%%%%%%%%%%%%%%%%%%%%%%%%%%%%%%%%%%

\subsection*{IGR\,J12171+7047}

The bright ($r\simeq12.61$ from the PanSTARRS data) galaxy NGC\,4250 (Fig. 2), which we consider as the most probable hard X-ray source, falls into the IBIS/INTEGRAL error circle. Moreover, previously the XRT/Swift telescope has already recorded soft X-ray emission from it. In the Swift source catalog this object is designated as Swift\,J1217.6+7047 (Evans et al. 2014).

We performed the object’s spectroscopy with the RTT-150 telescope. Narrow, redshifted H$\alpha$ (FWHM = 539 km s$^{-1}$, completely determined by the resolution of the instrument), [NII], [SII], and [OIII] lines are clearly distinguishable in the spectrum of the central part of the galaxy (Fig. 2). The measured redshift is $z=0.007\pm0.001$ and the corresponding photometric distance is $D_{L}=31.8 $ Mpc. This is in good agreement with the previous estimates of $D = 31-35$ Mpc (the NASA Extragalactic Database, NED). Its spectrum reveals a similarity both with the spectra of Seyfert 2 galaxies and with the spectra of LINER galaxies. To establish the true class of the object, we estimated the line flux ratios [NII]/H$\alpha$ = 1.27 and [SII]/H$\alpha$ = 0.88, the lower limit [OIII]/H$\beta > 2.7$, and used the well-known BPT diagram (Baldwin et al. 1981; Kewley et al. 2006). The values obtained correspond to the transition region between the LINER and Seyfert 2 galaxies. However, since the H$\beta$ line in the source’s spectrum is not statistically significant and since only a lower limit was obtained for the ratio [OIII]/H$\beta$, the investigated object is more likely a Seyfert 2 galaxy.

The total X-ray spectrum of IGR\,J12171+7047 and its putative counterpart Swift\,J1217.6+7047 is well fitted ($\chi^2=1.2$ per degree of freedom for 5 degrees of freedom) by a power law with a low energy cutoff due to photoabsorption. In this case, the slope is $\Gamma\simeq1.8$ and the absorption column density is $N_H\simeq10^{24}$ cm$^{-2}$. Given the distance to the object (see above), the hard X-ray luminosity of the source is $\sim 10^{42}$ erg s$^{-1}$ (Table 2). All of this suggests that we are dealing with a Seyfert 2 galaxy with a large absorption column density.

%================================================================
\begin{figure*}[t!]
\centering
\includegraphics[width=0.85\columnwidth,trim={0cm 0 0cm 0cm},clip]{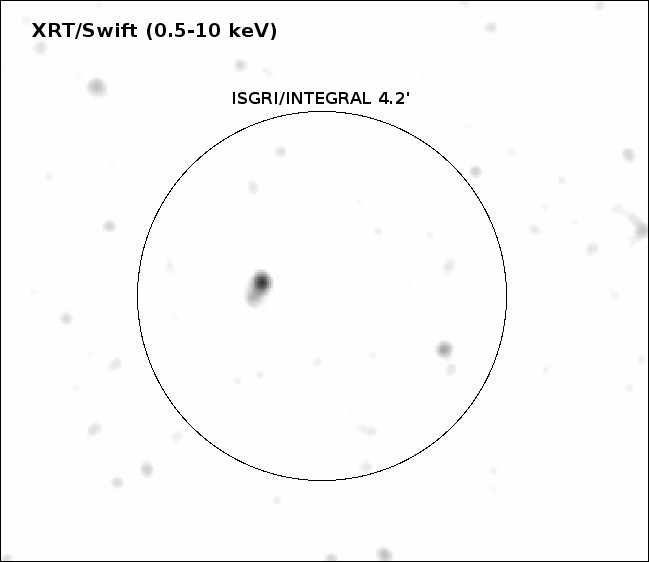}
\includegraphics[width=0.92\columnwidth,trim={-2cm 0 0cm 0cm},clip]{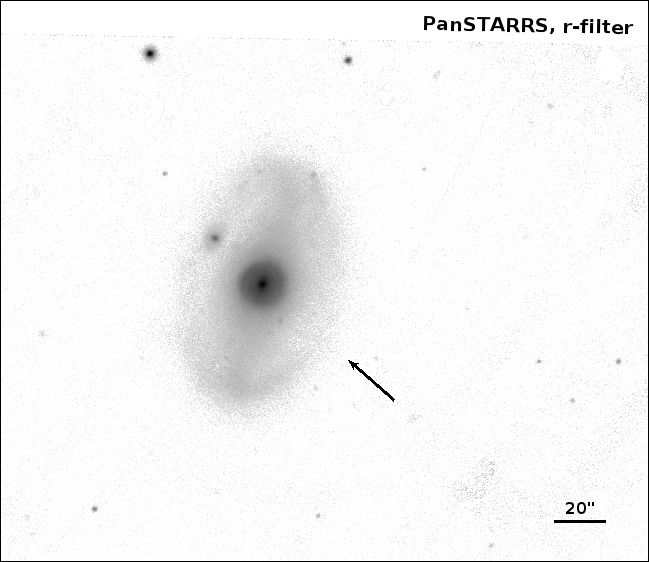}
\includegraphics[width=0.93\columnwidth,trim={0cm 7cm 0cm 3cm},clip]{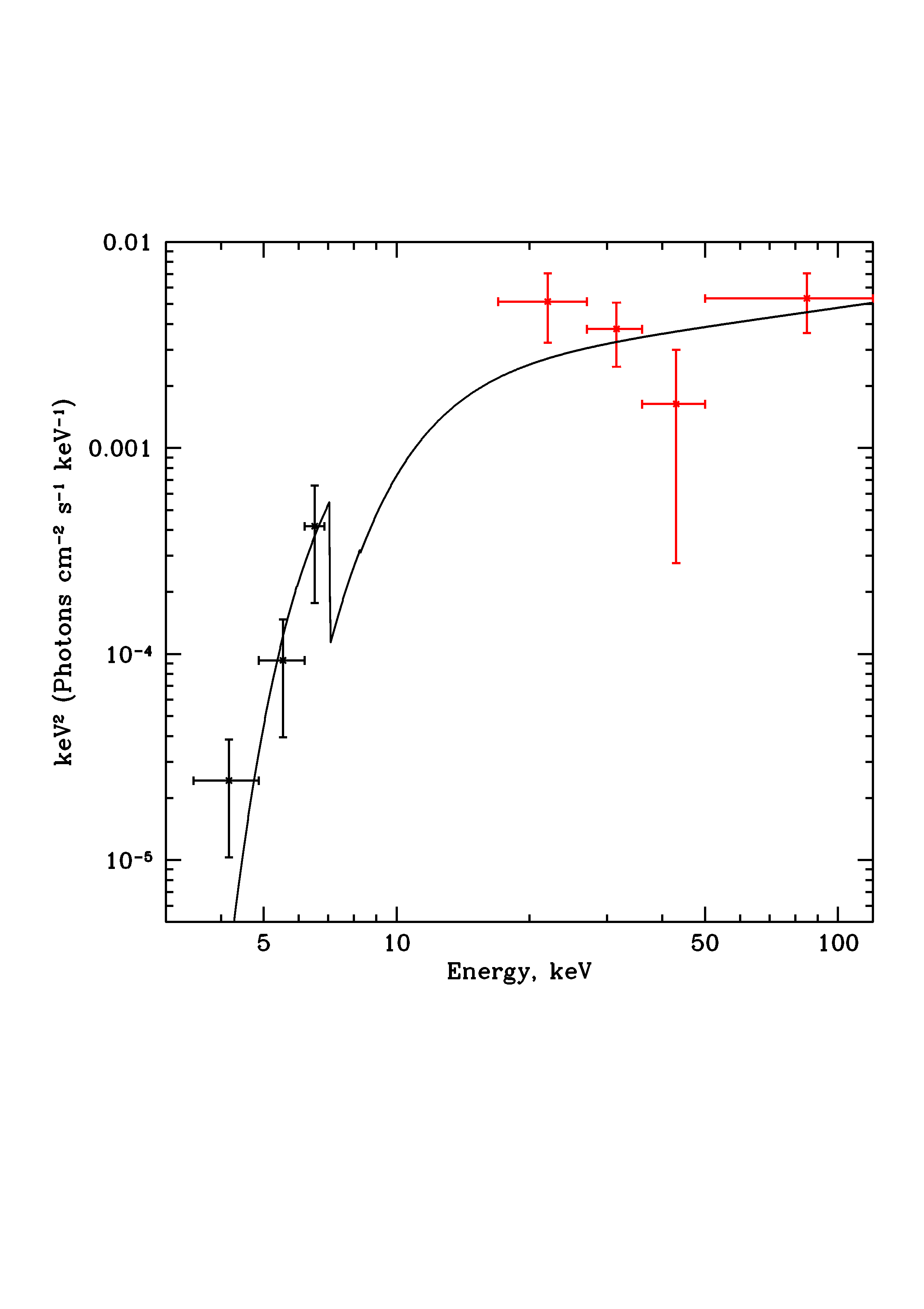}
\includegraphics[width=0.95\columnwidth,trim={0cm 0cm 0cm 3cm},clip]{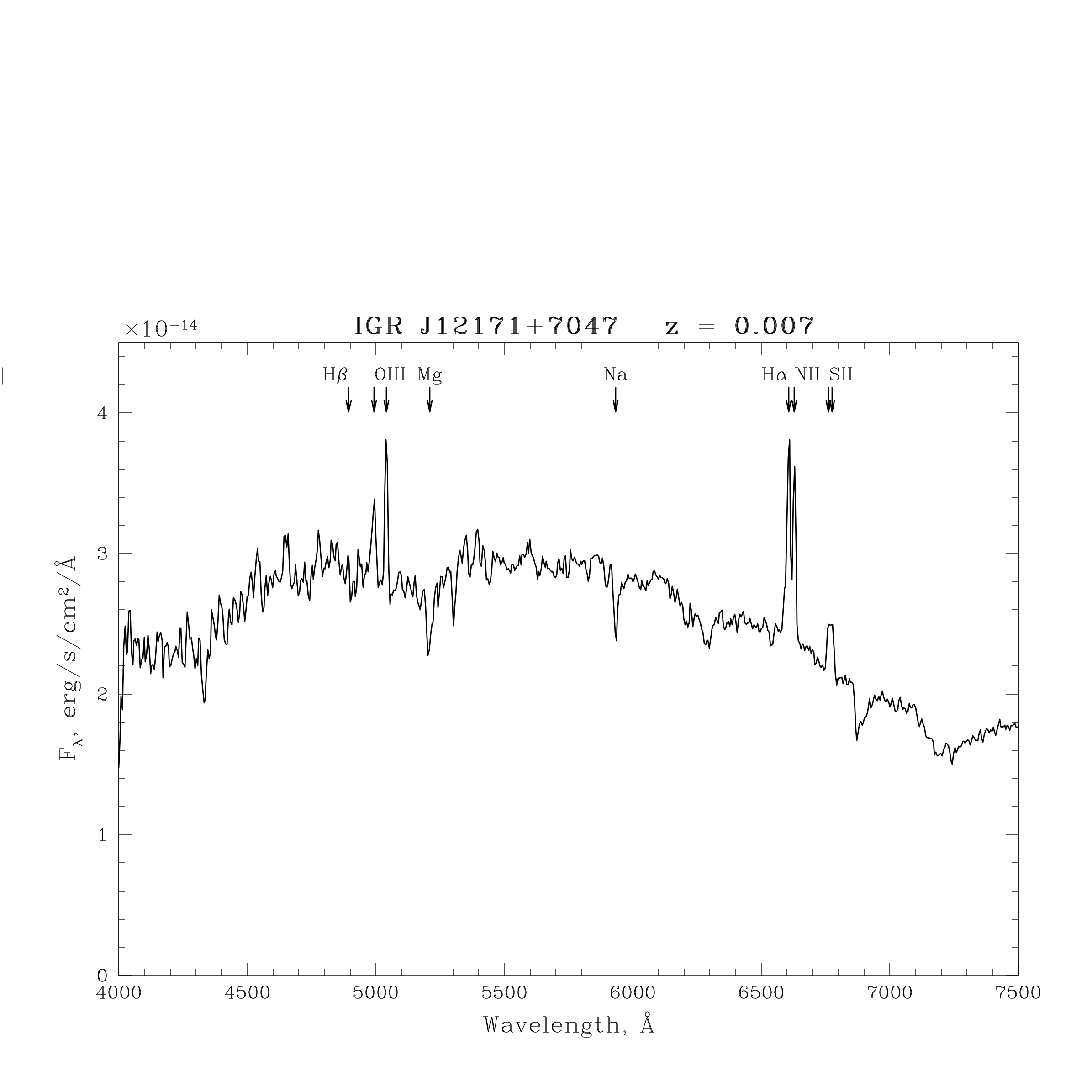}
\caption{Upper row: images of the sky region around IGR\,J12171+7047 from the XRT/Swift (left) and PanSTARRS (right) data. On the left panel the INTEGRAL error circle of the source is indicated by the solid circle. The arrow on the right panel marks the assumed hard X-ray source, the nearby galaxy NGC\,4250. Lower row: (left) the broadband
X-ray spectrum of the source from the XRT/Swift and IBIS/INTEGRAL data. The solid line indicates the fit to the spectrum by a power law with absorption; (right) the RTT-150 optical spectrum of the source. The most significant spectral lines are marked.}  \label{fig:IGR2}
\end{figure*}
%================================================================

%################################################################################################

%================================================================
\begin{figure*}[h!]
\centering
\includegraphics[width=0.85\columnwidth,trim={0cm 0 0cm 0cm},clip]{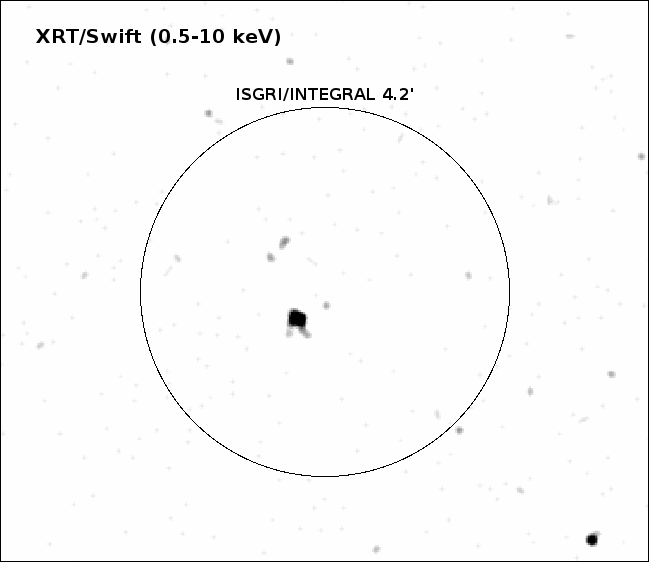}
\includegraphics[width=0.92\columnwidth,trim={-2cm 0cm 0cm 0cm},clip]{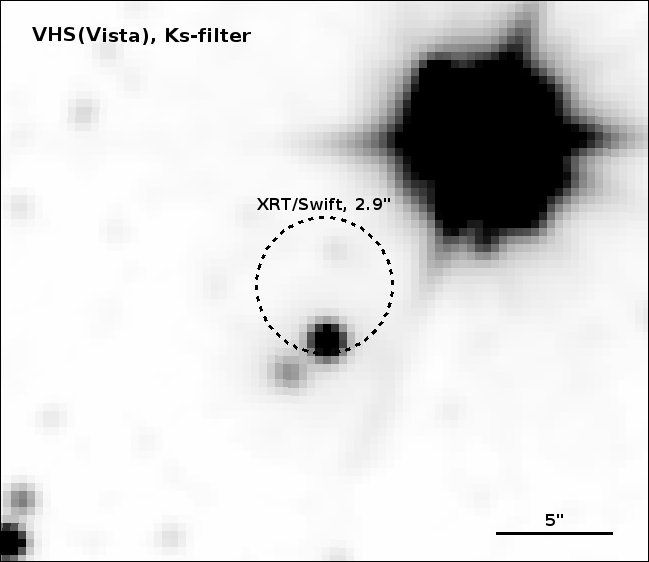}
\includegraphics[width=0.93\columnwidth,trim={0cm 7cm 0cm 3cm},clip]{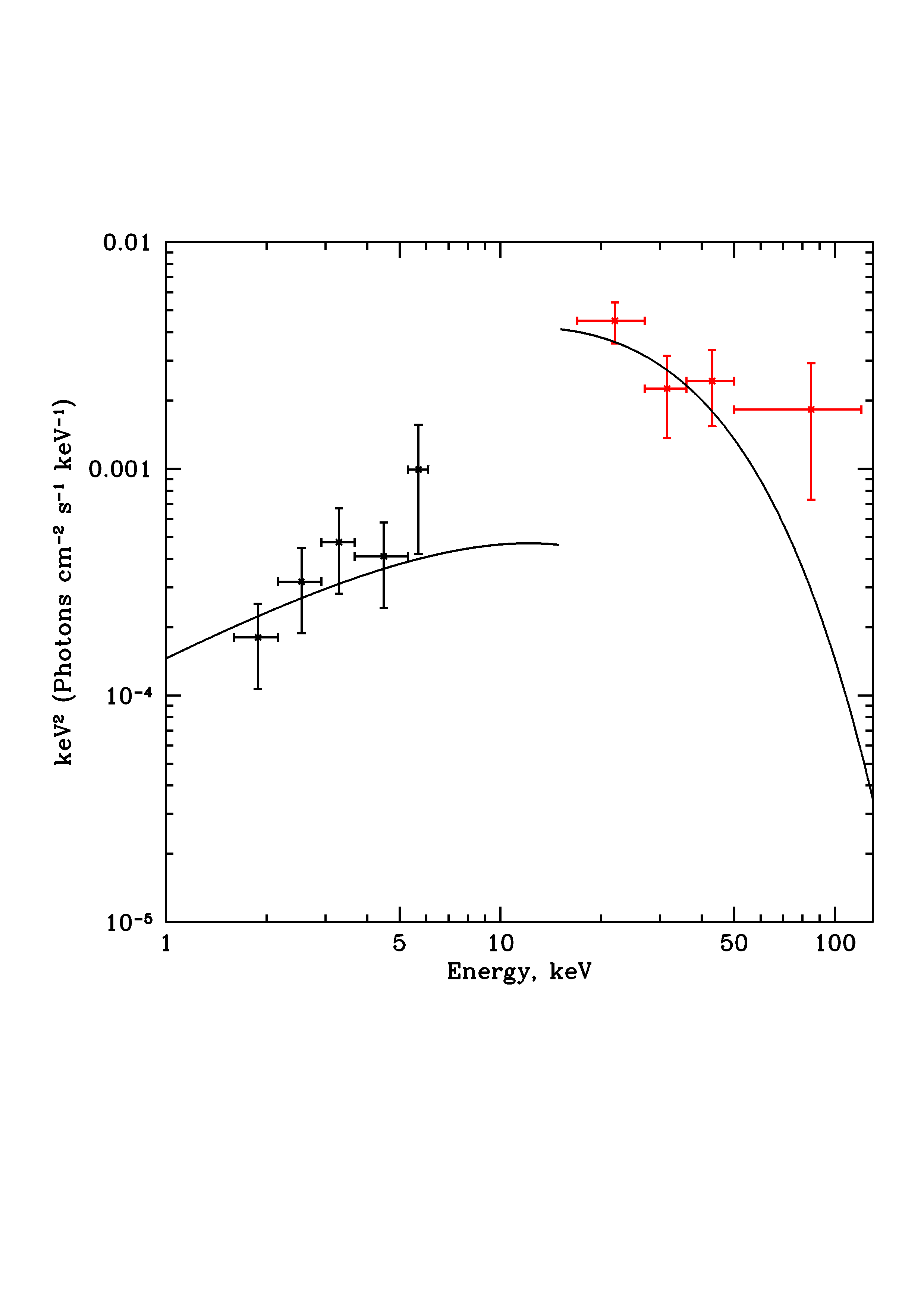}
\includegraphics[width=0.93\columnwidth,trim={0cm 7cm 0cm 3cm},clip]{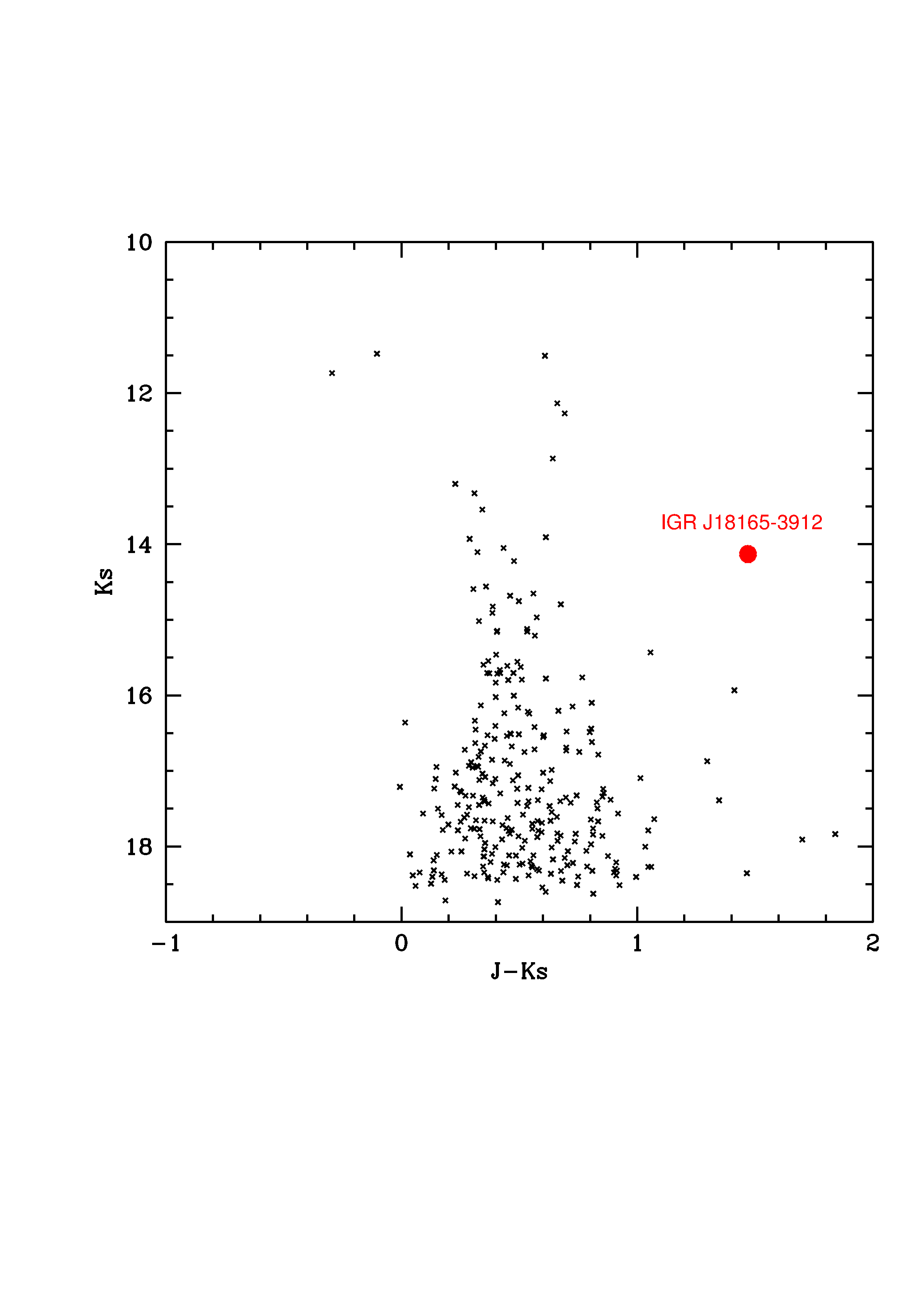}

\caption{Upper row: XRT/Swift (left) and VISTA (right) images of the sky region around IGR J18165--3912. The solid circle on the left panel indicates the INTEGRAL error circle of the source. The dashed circle on the right panel marks the XRT/Swift localization accuracy of the soft X-ray counterpart of the source. Lower row: the broadband X-ray spectrum
of IGR\,J18165--3912 from the XRT/Swift and IBIS/INTEGRAL data. The lines indicate the bremsstrahlung model with a plasma temperature $kT=19.1$ keV that best fits the experimental data points with different normalizations for the soft and
hard parts of the spectrum. The color–magnitude diagram constructed from the VHS sky survey data for all stars within $1'$of IGR\,J18165--3912. The position of the IR counterpart of the source is marked.
  }  \label{fig:IGR3}
\end{figure*}
%================================================================

%%%%%%%%%%%%%%%%%%%%%%%%%%%%%%%%%%%%%%%%%%%%%%%%                  3                      %%%%%%%%%%%%%%%%%%%%%%%%%%%%%%%%%%%%%%%%%%%%%%%%%%%%%%%%%%%%%%
\subsection*{IGR\,J18165--3912}

To improve the object’s localization, we requested the XRT/Swift observations (ObsID 00011480001, performed on July 16, 2019), based on which we managed to determine the soft X-ray counterpart of the source (Fig. 3) and to measure its X-ray flux
(Table 1). Having established an accurate position of the X-ray source, we were able to identify it in the optical and near-IR bands using data from the SkyMapper ($r=17.829\pm0.053$) and VHS (VISTA telescope) surveys of the European Southern Observatory (ESO). In addition, the object is recorded in the Gaia astrometric all-sky survey and, according to the estimates from Bailer-Jones et al. (2018), the distance to it is $6869^{+4300}_{-2763}$ pc. Thus, it is galactic in nature.

Using the VHS data, we constructed a color-magnitude diagram for all stars within $1'$ of the investigated object (Fig. 3). The IR counterpart of the  source is seen to lie slightly away from the main group of stars, which is typical for binary star systems with
accretion disks (and for AGNs). The dereddened (according to the VHS data, $E(J-Ks)\simeq0.05$) color of the object is $(J-Ks)=1.42\pm0.01$, which roughly
corresponds to the color of stars of spectral type M7 or later (Wegner 2014). The absolute magnitude of the IR counterpart of IGR\,J18165--3912, corrected for the extinction and the Gaia distance uncertainty, lies in the range $M_{Ks} = -1.29 \div 1.03$. For comparison,
$M_{Ks, M7III} = -7.08\pm0.73$ for a red M7 giant (Wegner 2007, 2014) and $M_{Ks, M7V} = 4.24\pm0.49$ for a main-sequence M7 star. Thus, we are probably dealing with the emission from an accretion disk around a compact object in a low-mass X-ray binary or a
cataclysmic variable (CV). 

The XRT/Swift and IBIS/INTEGRAL spectral data obtained (at different times) in the soft and hard X-ray energy bands agree poorly between themselves (Fig. 3), which may suggest a strong variability of the X-ray emission from the source. Making this assumption, the broadband spectrum of the source can be described, for example, by the plasma
bremsstrahlung model with a temperature $kT=19.1\pm5.9$ ($\chi^2 =0.52$  per degree of freedom for 10 degrees of freedom); this requires introducing a correction factor of 9.1 for the XRT/Swift data. Such a spectral shape is typical for the X-ray emission from
accretion columns on the poles of a magnetized white dwarf in CVs (polars and intermediate polars; Hailey et al. 2016). 

If IGR\,J18165--3912 is a low-mass X-ray binary, then its orbital period can be estimated based on a semiempirical relation from Revnivtsev et al. (2012). This formula relates the orbital period and luminosity of a binary in the X-ray (2-10 keV) and near-IR
bands. Taking into account the strong variability of IGR\,J18165--3912 and the uncertainty in its distance, we find that its orbital period should be 13.7--4400 days, which is quite atypical for low-mass X-ray binaries (the typical range is 0.01--10 days; Iben et al. 1997; Yan et al. 2010).

Based on the set of available data, we conclude that IGR\,J18165--3912 is most likely an intermediate polar -- a CV where matter from the companion star falls into the accretion disk and then onto the magnetic poles of the white dwarf. In particular, the object’s high X-ray luminosity ($\sim 10^{34}$--$10^{35}$ erg s$^{-1}$, given the significant uncertainty in the distance, see Table 2) provides evidence for this hypothesis. Moreover, IGR\,J18165--3912 is one of the most powerful (in X-rays) CVs discovered to date (Pretorius and Mukai 2014).

% VHS 2 stars;  
%CMC 15 r =17.138 0.458
%SkyMApper DR 1.1 17.829	0.053 
%Wolf et al 2018 https://ui.adsabs.harvard.edu/abs/2018PASA...35...10W/abstract

%###############################################################################################

%%%%%%%%%%%%%%%%%%%%%%%%%%%%%%%%%%%%%%%%%%%%%%%%                  4                     %%%%%%%%%%%%%%%%%%%%%%%%%%%%%%%%%%%%%%%%%%%%%%%%%%%%%%%%%%%%%%

%================================================================
\begin{figure*}[t!]
\centering
\includegraphics[width=0.90\columnwidth,trim={0cm -1cm 0cm 0cm},clip]{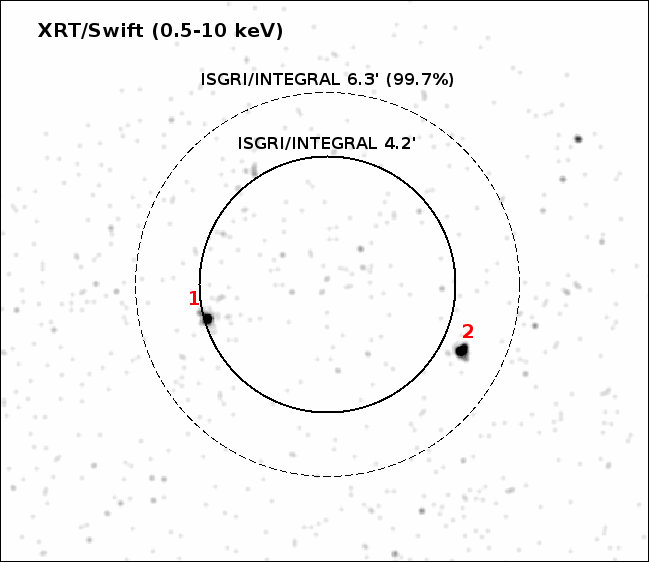}
\includegraphics[width=0.98\columnwidth,trim={-2cm -1cm 0cm 0cm},clip]{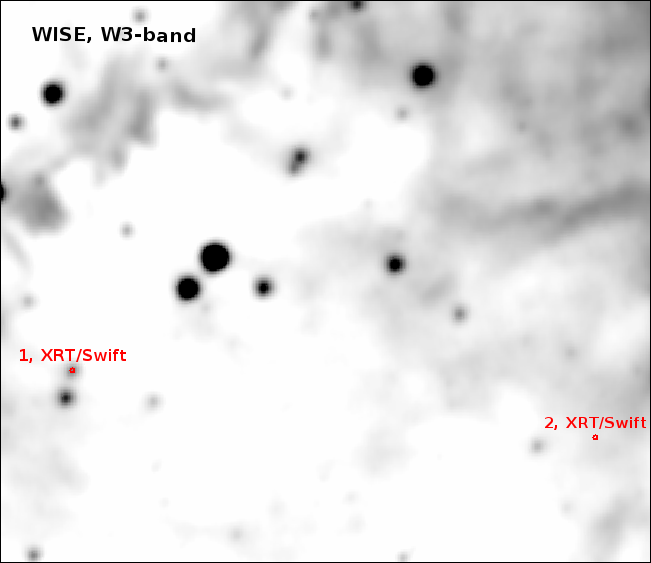}
\includegraphics[width=0.90\columnwidth,trim={0cm -1cm 0cm 0cm},clip]{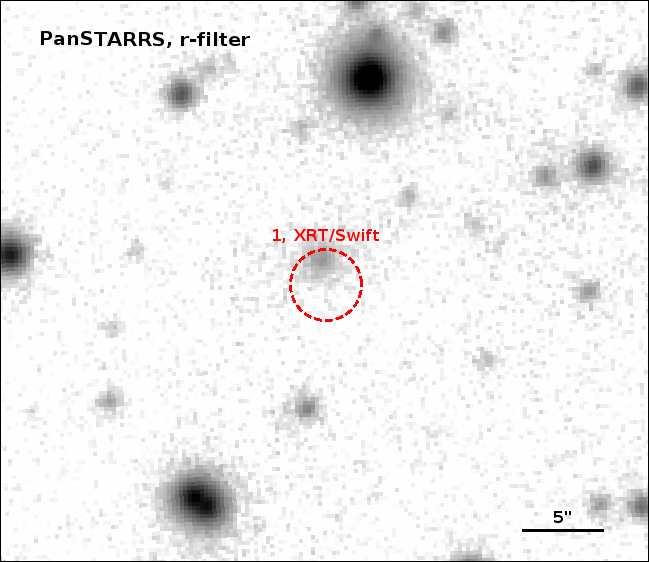}
\includegraphics[width=0.98\columnwidth,trim={-2cm -1cm 0cm 0cm},clip]{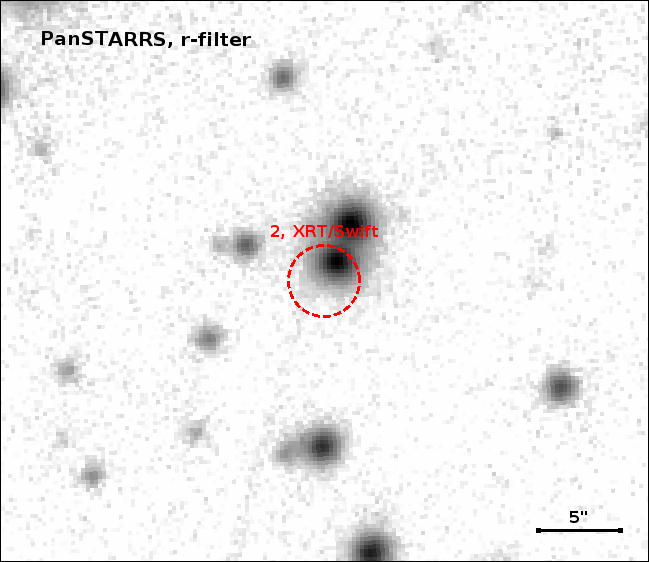}

\caption{XRT/Swift, WISE, and PanSTARRS images of the sky region around IGR J20596+4303. The solid and dashed circles indicate the INTEGRAL error circles of the source at 95.4\% and 99.7\% confidence, respectively. Numbers 1 and 2 mark the possible soft X-ray counterparts of the source and the optical and IR objects corresponding to them. The dashed circumferences reflect the XRT/Swift localization accuracy for the soft X-ray objects.}  \label{fig:IGR6}
\end{figure*}
%============================

%================================================================
\begin{figure*}[t!]

\centering
\includegraphics[width=0.95\columnwidth,trim={0cm 7cm 0cm 5cm},clip]{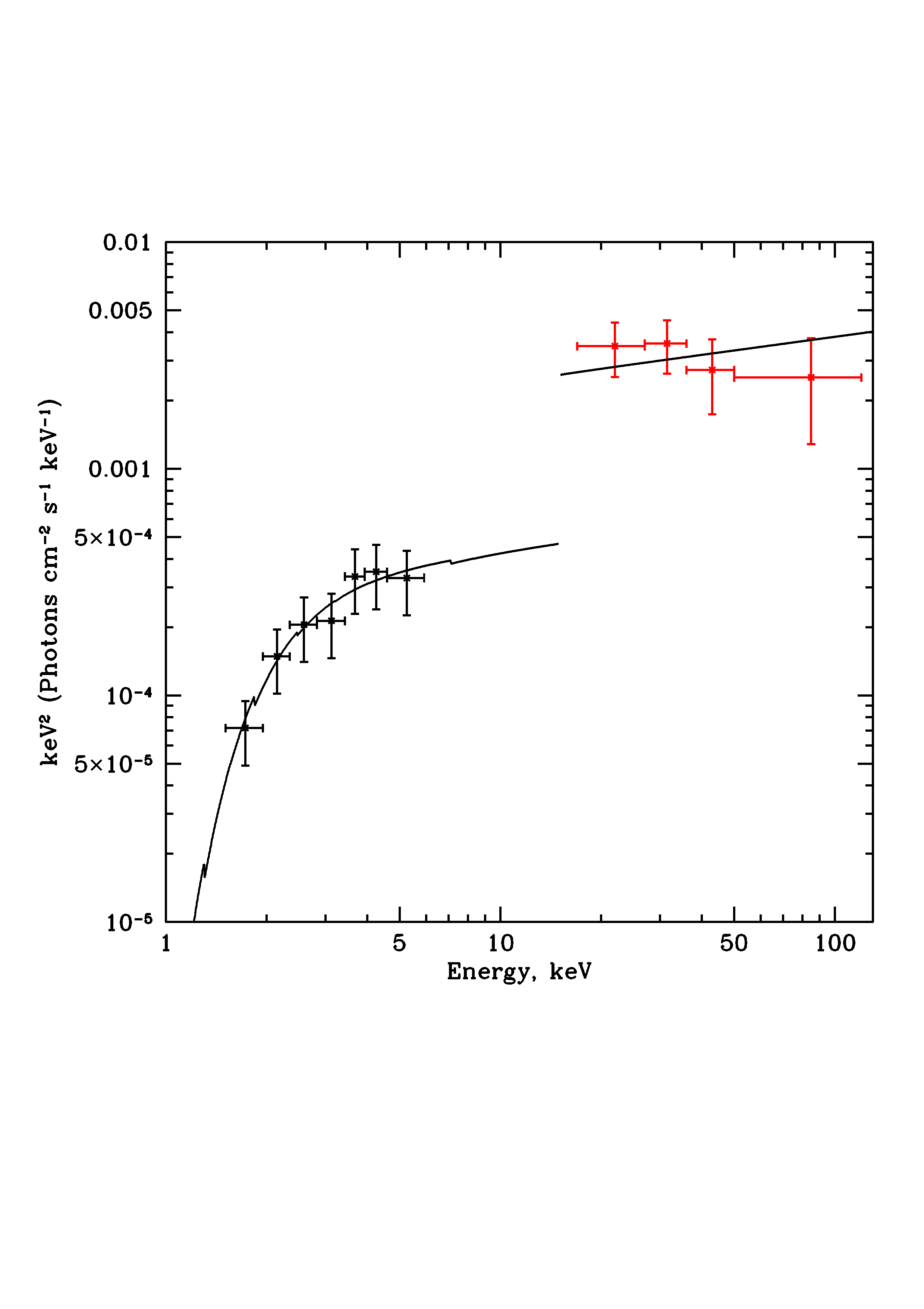}
\includegraphics[width=0.95\columnwidth,trim={0cm 7cm 0cm 5cm},clip]{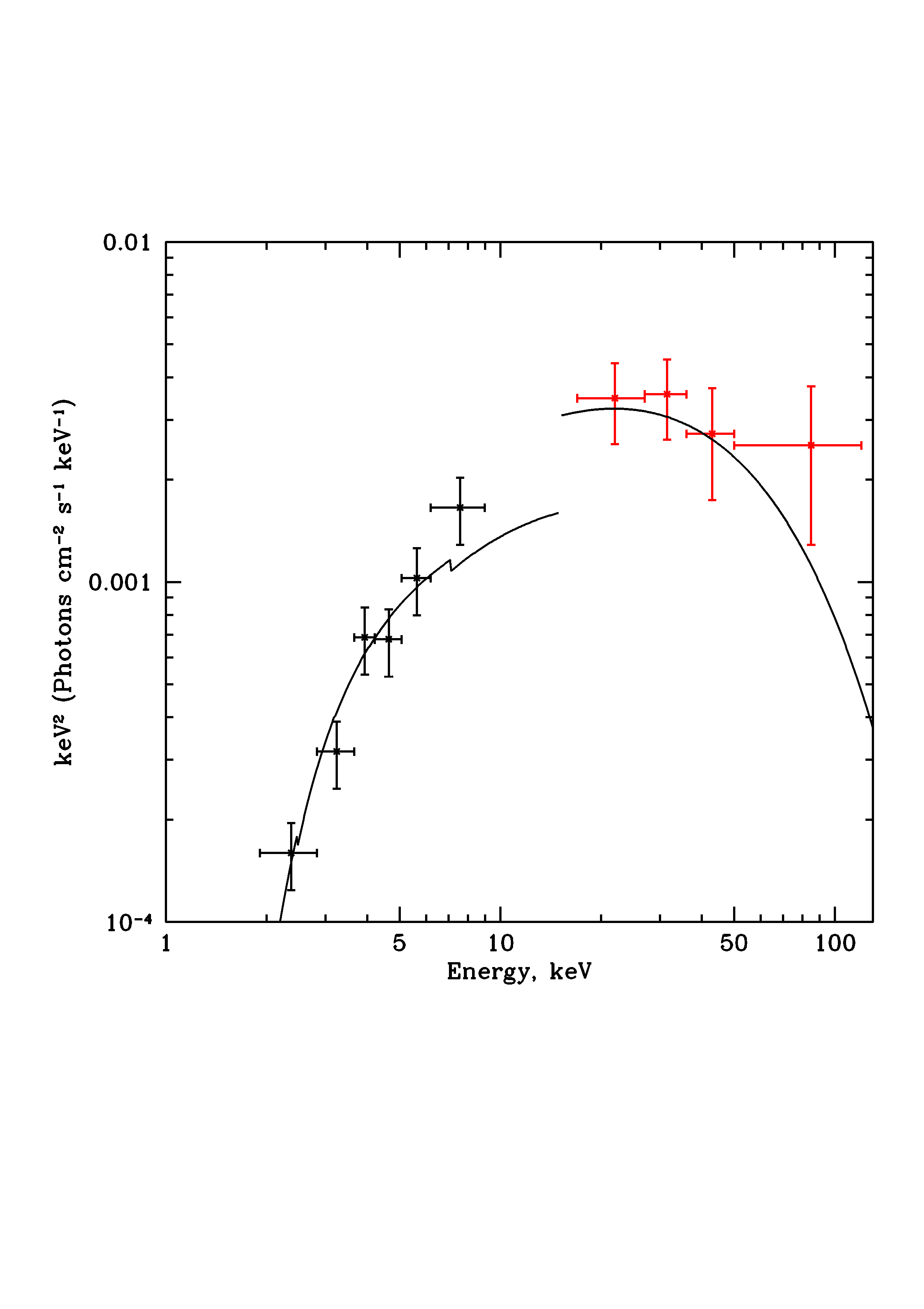}
\includegraphics[width=0.95\columnwidth,trim={0cm 0 0cm 3cm},clip]{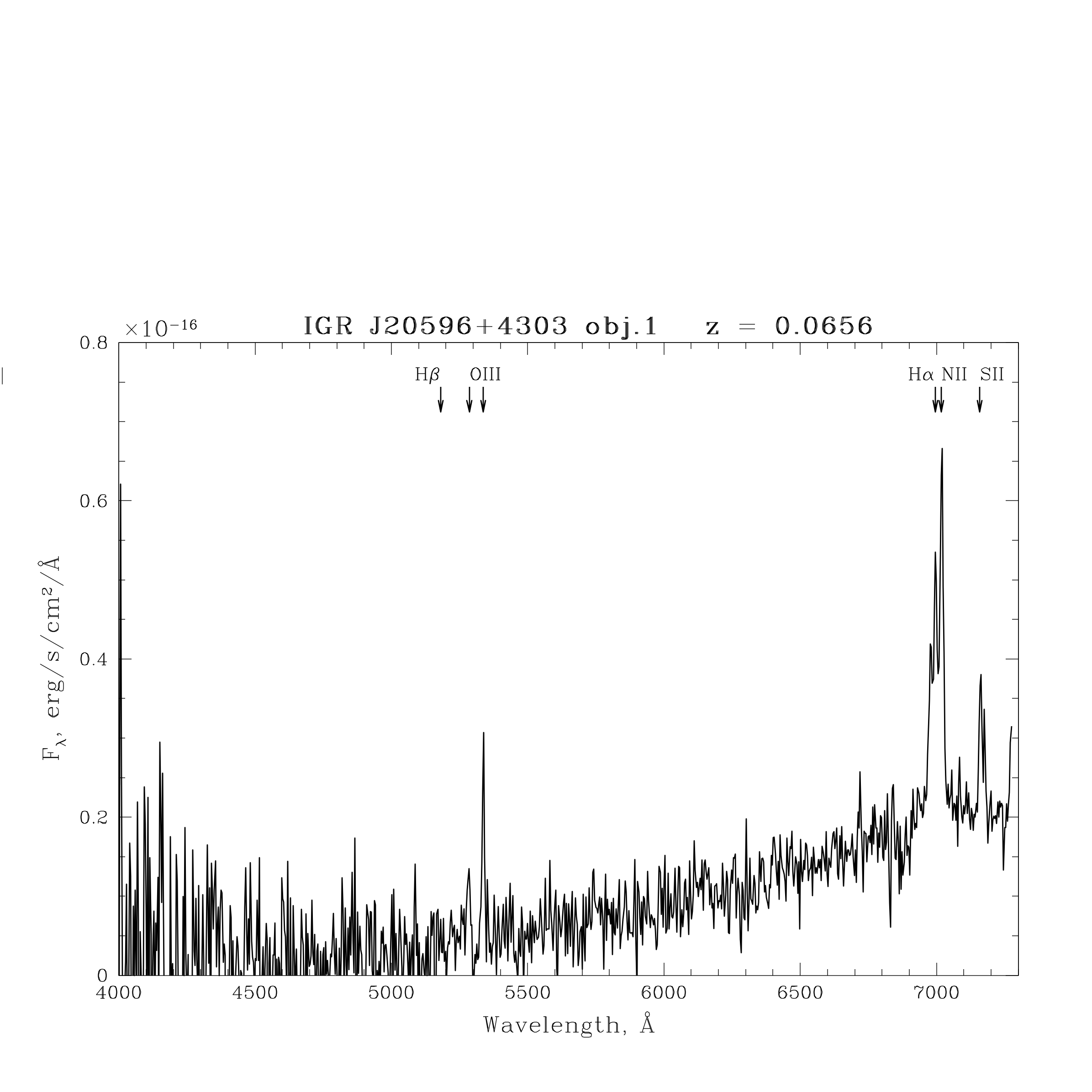}
\includegraphics[width=0.95\columnwidth,trim={0cm 0cm 0cm 3cm},clip]{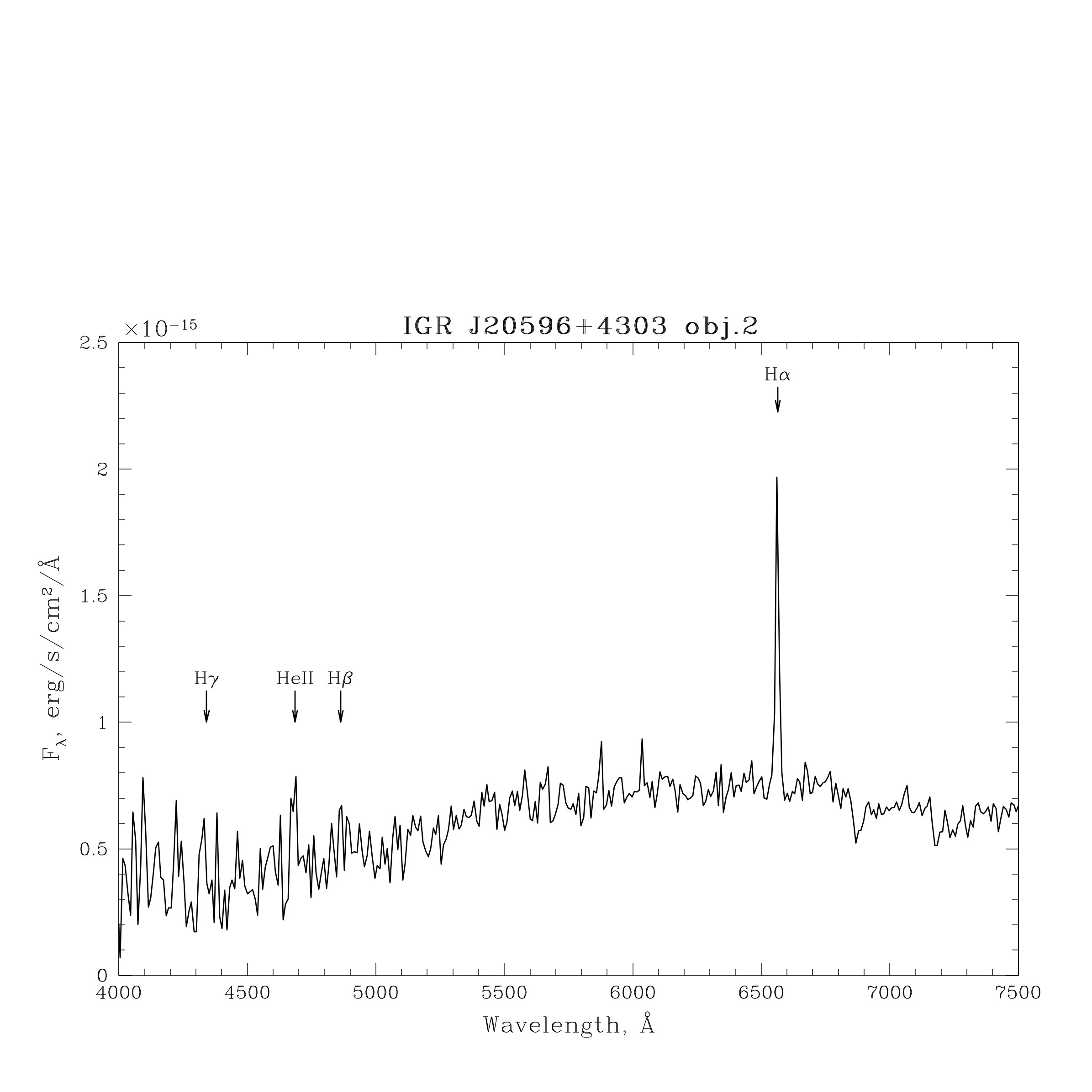}

\caption{Upper rows: the X-ray spectra of the assumed soft X-ray counterparts of IGR\,J20596+4303 (black dots, SWIFT\,J2059.6+4301B/1SXPS\,J210000.9+430208 (left) and SWIFT\,J2059.6+4301A/1SXPS\,J205915.6+430105 (right)) combined with the INTEGRAL hard X-ray spectrum of the source (red dots). The lines indicate the models that best
fit these data points (see the text). Lower row: the corresponding optical spectra of the putative counterparts obtained with the following telescopes: AZT-33IK (SWIFT\,J2059.6+4301B, left) and RTT-150 (SWIFT\,J2059.6+4301A, right). The most
significant spectral lines are marked.}  \label{fig:IGR6_1}
\end{figure*}
%============================

%================================================================
\begin{figure*}

\centering
\includegraphics[width=1.2\columnwidth,trim={0cm 7cm 0cm 5cm},clip]{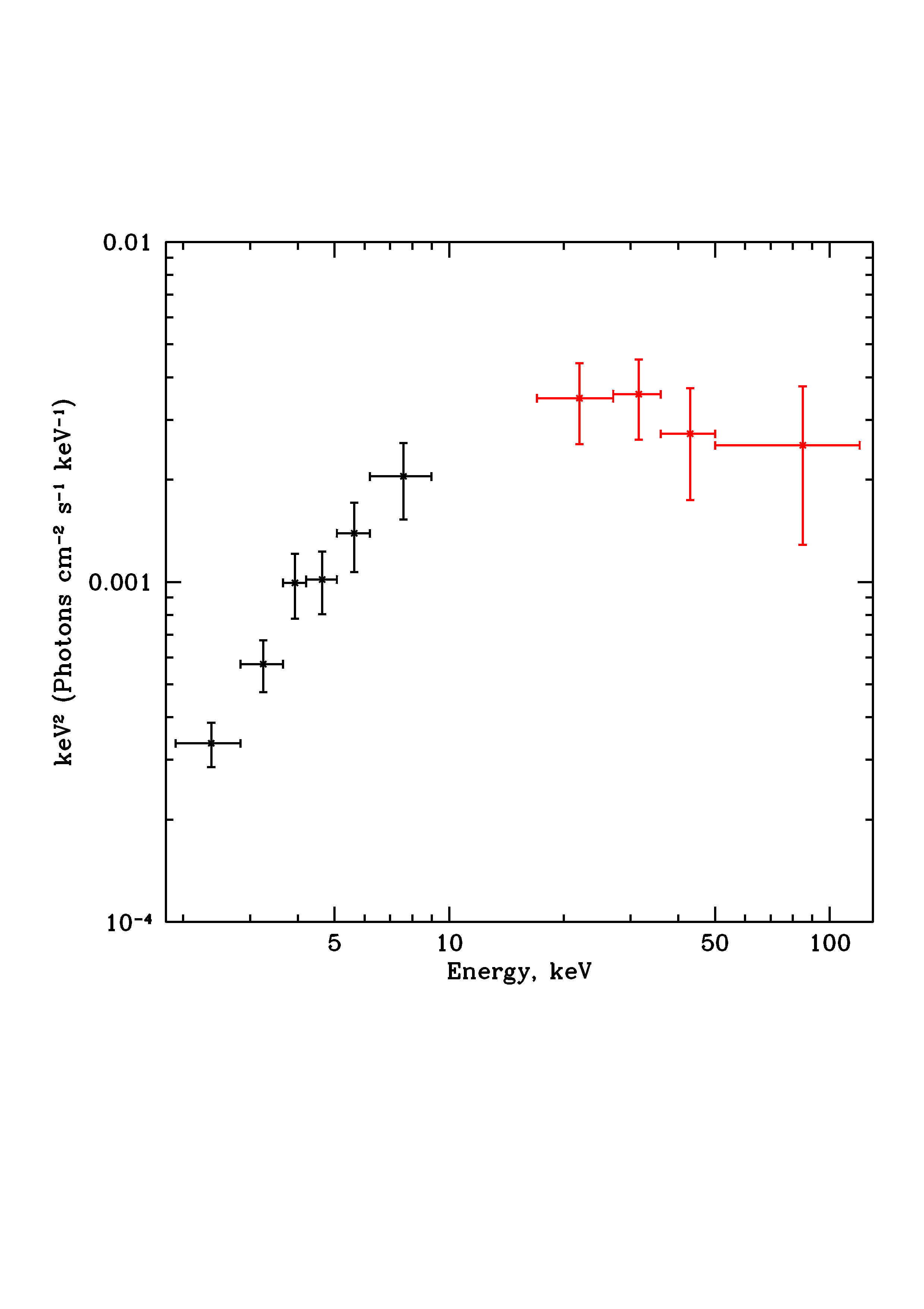}

\caption{The total X-ray spectrum of the two supposed soft X-ray counterparts of IGR\,J20596+4303 (SWIFT\,J2059.6+4301B/1SXPS\,J210000.9+430208 plus SWIFT\,J2059.6+4301A/1SXPS\,J205915.6+430105, black dots) combined with the INTEGRAL hard X-ray spectrum of the object (red dots).
}  \label{fig:IGR6_2}
\end{figure*}
%============================

\subsection*{IGR\,J20596+4303}

According to the Swift data, there are two X-ray sources (Fig. 4) known as SWIFT\,J2059.6+4301A/1SXPS\,J205915.6+430105\footnote{https://swift.gsfc.nasa.gov/results/bs105mon/1095} and SWIFT\,J2059.6+4301B/1SXPS J210000.9+430208\footnote{https://swift.gsfc.nasa.gov/results/bs70mon
/SWIFT\_J2059.6p4301B}  in the immediate vicinity of this object (Baumgartner et al. 2013).
SWIFT\,J2059.6+4301B is slightly closer ($4.1'$) to the position measured by the IBIS/NTEGRAL telescope, falling into the 2$\sigma$ error circle. SWIFT\,J2059.6+4301A is $4.7'$ away from
IGR,J20596+4303 and falls into the 3$\sigma$ error circle with a radius of $6.3'$.

An accurate localization of SWIFT\,J2059.6+4301B based on the XRT data
(Table 2) allows its optical counterpart to be unambiguously determined. This is an object with  optical/IR magnitudes $r\simeq20.37$, $W1=12.63\pm0.03$  (from the PanSTARRS and WISE data, respectively) and with infrared colors ($W1-W2 \approx 0.73$)
typical for AGNs (Stern et al. 2012). Previously, this object has already been noted as an AGN candidate (Edelson and Malkan 2012). Our AZT-33IK spectrum shows that this is a Seyfert 2 galaxy (FWHM$_{H\alpha} = 546.6$ km s$^{-1}$, completely determined
by the resolution of the instrument) at redshift $z=0.0656\pm0.0010$ (Fig. 5 on the left).

The hard X-ray flux measured with the IBIS/INTEGRAL instrument is higher than the soft
X-ray flux from SWIFT\,J2059.6+4301B by an order of magnitude (Fig. 5). If this difference is attributed to the object’s variability, then the total spectrum of the pair IGR\,J20596+4303/SWIFT\,J2059.6+4301B
can be described ($\chi^2=0.31$ per degree of freedom for 8 degrees of freedom) by a power law ($\Gamma\simeq1.8$) with absorption $N_H=2.6\pm0.7\times10^{22}$ cm$^{-2}$, typical values for Seyfert 2 galaxies. This requires introducing an additional normalization factor of 5.6 for the XRT/Swift data.

As regards the other possible X-ray counterpart, SWIFT\,J2059.6+4301A, according to the optical and IR survey data, a fairly bright optical object ($r\simeq17.53$), whose IR colors ($W1-W2\sim0$) point to its stellar nature, falls into the error circle of this source.
According to the Gaia data (Bailer-Jones et al. 2018), the distance to the object is $2068^{+692}_{-432}$ pc. An unshifted narrow H$\alpha$ emission line (Fig. 5 on the right), pointing to the emission from an accretion disk in a binary system (for example, a CV), is identified in the RTT-150 spectrum.

The total spectrum of the pair IGR\,J20596+4303/ SWIFT\,J2059.6+4301A can be described ($\chi^2=0.89$ per degree of freedom for 6 degrees of freedom)
by the bremsstrahlung model with a plasma temperature $kT=33\pm15$  and absorption 
$N_H=5.19\pm1.24\times10^{22} $  cm$^{-2}$, typical for magnetic CVs.
However, this requires introducing an additional normalization factor for the XRT/Swift data, i.e., again admitting a strong variability of the object. The measured high X-ray luminosity of the source (Table 2), $\gtrsim 10^{33}$ erg s$^{-1}$, suggests that this is an intermediate polar.

Taking into account what has been said above, the hard X-ray source IGR\,J20596+4303 is apparently a superposition of both objects considered,
SWIFT\,J2059.6+4301A and SWIFT\,J2059.6+4301B, the first of which is a CV and the second is an AGN. To better demonstrate this, we constructed the total soft X-ray spectrum of SWIFT\,J2059.6+4301A and SWIFT\,J2059.6+4301 from the XRT/Swift data
and compared it with the spectrum of IGR\,J20596+4303 measured by INTEGRAL at higher energies(Fig. 6).

Note also that the results obtained by us separately
for SWIFT\,J2059.6+4301A and SWIFT\,J2059.6+4301B agree well with those presented in Marchesini et al. (2019).

\section*{CONCLUSIONS}

We made optical identifications of four hard X-ray sources first detected during a deep extragalactic survey (Mereminskiy et al. 2016) and an extended Galactic plane survey ($|b|<17^\circ$, Krivonos et al. 2017) with the INTEGRAL observatory. We
showed that two of them (IGR\,J11079+7106 and IGR\,J12171+7047) are nearby Seyfert 1 and 2 galaxies, respectively. The redshifts were measured for them, with the object IGR\,J12171+7047 being characterized by a large absorption column density
($N_H\simeq10^{24}$ cm$^{-2}$). The third object (IGR\,J18165--3912) is apparently a cataclysmic variable (intermediate
polar)with a very high luminosity ($\gtrsim 10^{34}$ erg s$^{-1}$.
Finally, the fourth object (IGR\,J20596+4303) is most probably a superposition of two X-ray sources of comparable brightness, one of which is a Seyfert 2 galaxy and the other one is a cataclysmic variable (most likely an intermediate polar).

%\begin{landscape}

\begin{table*}[t]
\centering
\footnotesize{
 \begin{flushleft}  
   \caption{Optical and IR counterparts of the hard X-ray sources}
 \end{flushleft}     
 
   \resizebox{18.2cm}{!} {   
   \begin{tabular}{c|c|c|c|c|c|c|c|c}
     \hline
     \hline
                               &       &      &   &   &  & & &\\
      Name        & RA (J2000) & Dec (J2000) &  Redshift/distance & $log L_{2-10}$  & $log L_{17-60}$ & r {\tiny (PanSTARRS)} & Class of object &Notes\\
              &        &       &       &  &  & & &\\

     \hline
\rule{0cm}{0.5cm}
     IGR\,J11079+7106    &  11$^h$07$^m$47$^s$.86 &  71\dgr 05\arcmin 32\arcsec.44 &    0.059 /  272.6 Mpc & $43.30^{+0.08}_{-0.08}$   &  $43.81^{+0.04}_{-0.04}$  & $16.97\pm0.01$  & Seyfert 1 &2MASX\,J11074777+7105326\\ [0.25cm]
     
     \rule{0cm}{0.5cm}
     IGR\,J12171+7047   &  12$^h$17$^m$26$^s$.22 &  +70\dgr 48\arcmin 09\arcsec.00 & 0.007 / 31.6 Mpc  &$40.16^{+0.15}_{-0.15}$ & $42.08^{+0.06}_{-0.06}$ & $\simeq12.61$ &Seyfert 2 &NGC\,4250  \\ [0.25cm]
     
     \rule{0cm}{0.5cm}
     IGR\,J18165--3912    &  18$^h$16$^m$35$^s$.941 & -39\dgr 12\arcmin 46\arcsec.21 & 0 / $6.9^{+4.3}_{-2.8}$ kpc & $33.94^{+0.44}_{-0.51}$ & $34.54^{+0.44}_{-0.47}$ & $17.83\pm0.05$ & CV & 2MASS\,J18163594--3912464\\ [0.25cm]

     \rule{0cm}{0.5cm} 
     IGR\,J20596+4303$_1$	 &  21$^h$00$^m$01$^s$.00 &  +43\dgr 02\arcmin 10\arcsec.97 & 0.0656 / 304.6 Mpc &$43.14^*$ & $<43.80^*$& $20.37\pm0.06$ &Seyfert 2  & WISE\,J210000.99+430210.9\\  [0.25cm]
     
     -$_2$	 &  20$^h$59$^m$15$^s$.70 &  +43\dgr 01\arcmin 07\arcsec.27 & 0 / $2.1^{+0.7}_{-0.4}$ kpc &$33.21^*$ & $<33.47^*$ & $17.53\pm0.09$ & CV  & UGPS\,J205915.69+430107.2\\  [0.25cm]

     \hline
    \end{tabular}
    }
    
  \begin{flushleft}  
    *  The flux from IGR\,J20596+4303 in the 17–60 keV energy band is presumably a superposition of the fluxes from two separate sources, SWIFT\,J2059.6+4301A and SWIFT\,J2059.6+4301B. However, given the possible variability of the sources, it is fairly difficult to
estimate the contribution of each of them to the resulting flux.
\end{flushleft}
    }    
    
\end{table*}

%\end{landscape}

\bigskip
\section*{ACKNOWLEDGMENTS}
This work was supported by RSF 19-12-00396 grant. We thank the TUBITAK National Observatory (Turkey), the Space Research Institute of the Russian
Academy of Sciences, and the Kazan State University
for the support in using the Russian--Turkish 1.5-m telescope
(RTT-150). The results at the AZT-33IK telescope
were obtained using the equipment of the Angara Center
http://ckp-rf.ru/ckp/3056/ within the Basic Financing
Program FNI II.16. We used the VHS data obtained with
the VISTA/ESO telescope (Paranal Observatory) within
Program 179.A-2010 (PI: McMahon). This publication
also uses data from the WISE observatory, which is a
joint project of the California University, Los Angeles,
and the Jet Propulsion Laboratory/California Institute of
Technology financed by National Aeronautics and Space
Administration.

%%%%%%%%%%%%%%%%%%%%%%%%%%%

\section*{REFERENCES}

1. V. Afanasiev, S. Dodonov, V. Amirkhanyan, and A. Moiseev, Astrophys. Bull. {\bf 71}, 479 (2016).\\

2. C. A. L. Bailer-Jones, J. Rybizki, M. Fouesneau, G. Mantelet, and R. Andrae, Astrophys. J. Lett. {\bf 156}, 58B (2018). \\

3. J. A. Baldwin, M. M. Phillips, and R. Terlevich, Publ. Astron. Soc. Pacif. {\bf 93}, 5 (1981). \\

4. W. H. Baumgartner, J. Tueller, C. B. Markwardt, G. K. Skinner, S. Barthelmy, et al., Astron. Astrophys. Suppl. Ser. {\bf 207}, 19 (2013). \\

5. I. Bikmaev, M. Revnivtsev, R. Burenin, and R. Syunyaev, Astron. Lett. {\bf 32}, 588 (2006).\\

6. I. Bikmaev, R. Burenin, M. Revnivtsev, S. Yu. Sazonov, R. A. Sunyaev, M. N. Pavlinsky, and N. A. Sakhibullin, Astron. Lett. {\bf 34}, 653 (2008).\\

7. A. J. Bird, A. Bazzano, A. Malizia, M. Fiocchi, V. Sguera, L. Bassani, A. B. Hill, P. Ubertini, and C. Winkler, Astrophys. J. Suppl. Ser. {\bf 223}, 10 (2016).\\

8. R. Burenin, A. Meshcheryakov, M. Revnivtsev, S. Yu. Sazonov, I. F. Bikmaev, M. N. Pavlinsky, and R. A. Sunyaev, Astron. Lett. {\bf 34}, 367 (2008).\\

9. R.Burenin, I.Bikmaev, M.Revnivtsev, J.A.Tomsick, S. Yu. Sazonov, M. N. Pavlinsky, and R. A. Sunyaev, Astron. Lett. {\bf 35}, 71 (2009).\\

10. R. A. Burenin, A. L. Amvrosov, M. V. Eselevich, V. M. Grigor’ev, V. A. Aref’ev, V. S. Vorob’ev, A. A. Lutovinov, M. G. Revnivtsev, S. Yu. Sazonov, A. Yu. Tkachenko, G. A. Khorunzhev, A. L. Yaskovich, and M. N. Pavlinsky, Astron. Lett. {\bf 42}, 295 (2016).\\

11. E.Churazov, R.Sunyaev, S.Sazonov, M.Revnivtsev, and D. Varshalovich, Mon. Not. R. Astron. Soc. {\bf 357}, 1377 (2005).\\

12. E. Churazov, R. Sunyaev, J. Isern, et al., Nature (London, U.K.) {\bf 512}, 406 (2014).\\

13. R. Edelson and M. Malkan, Astrophys. J. {\bf 751}, 52 (2012). \\

14. P. A. Evans, A. P. Beardmore, K. L. Page, J. P. Osborne, P. T. O’Brien, R. Willingale, R. L. C. Starling, D. N. Burrows, et al., Mon. Not. R. Astron. Soc. {\bf 397}, 1177 (2009).\\

15. P. A. Evans, J. P. Osborne, A. P. Beardmore, K. L. Page, R. Willingale, C. J. Mountford, C. Pagani, D. N. Burrows, et al., Astrophys. J. Suppl. Ser. {\bf 210}, 24 (2014). \\

16. M.R.Goad, L.G.Tyler, A.P.Beardmore, P.A.Evans, S. R. Rosen, J. P. Osborne, R. L. C. Starling, F. E. Marshall, et al., Astron. Astrophys. {\bf 476}, 1401 (2007). \\

17. S. A. Grebenev, A. A. Lutovinov, S. S. Tsygankov, and I. A. Mereminskiy, Mon. Not. R. Astron. Soc. {\bf 428}, 50 (2013). \\

18. C. J. Hailey, K. Mori, K. Perez, A. M. Canipe, J. Hong, et al., Astrophys. J. {\bf 826}, 160 (2016). \\

19. I. Iben, Jr., A. V. Tutukov, and A. V. Fedorova, Astrophys. J. {\bf 486}, 955 (1997). \\

20. S.F.Kamus, S.A.Denisenko, N.A.Lipin, V.I.Tergoev, P. G. Papushev, S. A. Druzhinin, Yu. S. Karavaev, and Yu. M. Palachev, J. Opt. Technol. {\bf 69}, 674 (2002). \\

21. D. I. Karasev, A. A. Lutovinov, and R. A. Burenin, Mon. Not. R. Astron. Soc. Lett. {\bf 409}, L69 (2010).\\

22. D. I. Karasev, A. A. Lutovinov, A. Yu. Tkachenko, G. A. Khorunzhev, R. A. Krivonos, P. S. Medvedev, M. N. Pavlinsky, R. A. Burenin, and M. V. Eselevich, Astron. Lett. {\bf 44}, 522 (2018). \\

23. L. J. Kewley, B. Groves, G. Kauffmann, and T. Heckman, Mon. Not. R. Astron. Soc. {\bf 372}, 961 (2006). \\

24. R. Krivonos, M. Revnivtsev, A. Lutovinov, S. Sazonov, E. Churazov, and R. Sunyaev, Astron. Astrophys. {\bf 475}, 775 (2007).\\

25. R. Krivonos, M. Revnivtsev, S. Tsygankov, et al., Astron. Astrophys. {\bf 519}, A107 (2010). \\

26. R.Krivonos, S.Tsygankov, A.Lutovinov, M.Revnivtsev, E. Churazov, and R. Sunyaev, Astron. Astrophys. {\bf 545}, 7 (2012). \\

27. R. Krivonos, S. Tsygankov, I. Mereminskiy, A. Lutovinov, S. Sazonov, and R. Sunyaev, Mon. Not. R. Astron. Soc. {\bf 470}, 512 (2017).\\

28. A.Lutovinov, R.Burenin, S.Sazonov, M.Revnivtsev, A. Moiseev, and S. Dodonov, Astron. Telegram 2759, 1 (2010). \\

29. A.Malizia,L.Bassani,V.Sguera,etal.,Mon.Not.R. Astron. Soc. {\bf 408}, 975 (2010). \\

30. E. J. Marchesini, N. Masetti, E. Palazzi, V. Chavushyan, E. Jimenez-Bailon, et al., Astrophys. Space Sci. {\bf 364}, 153 (2019). \\

31. N. Masetti, R. Landi, M. Pretorius, et al., Astron. Astrophys. {\bf 470}, 331 (2007). \\

32. N.Masetti, P.Parisi, E.Palazzi, etal., Astron. Astrophys. {\bf 519}, 96 (2010). \\

33. I. A. Mereminskiy, R. A. Krivonos, A. A. Lutovinov, S. Yu. Sazonov, M. G. Revnivtsev, and R. A. Sunyaev, Mon. Not. R. Astron. Soc. {\bf 459}, 140 (2016). \\

34. S.V.Molkov, A.M.Cherepashchuk, A.A.Lutovinov, M. G. Revnivtsev, K. A. Postnov, and R. A. Sunyaev, Astron. Lett. {\bf 30}, 382 (2004). \\

35. Planck Collab., Astron. Astrophys. {\bf 594}, 63 (2016). \\

36. M. L. Pretorius and K. Mukai, Mon. Not. R. Astron. Soc. {\bf 442}, 2580 (2014). \\

37. M. Revnivtsev, R. Sunyaev, D. Varshalovich, V. Zheleznyakov, A. Cherepashchuk, A. Lutovinov, E. Churazov, S. Grebenev, and M. Gilfanov, Astron. Lett. {\bf 30}, 534 (2004). \\

38. M. G. Revnivtsev, S. Yu. Sazonov, S. V. Molkov, A. A. Lutovinov, E. M. Churazov, and R. A. Sunyaev, Astron. Lett. {\bf 32}, 145 (2006). \\

39. M. G. Revnivtsev, I. Yu. Zolotukhin, and A. V. Meshcheryakov, Mon. Not. R. Astron. Soc.
{\bf 421}, 2846 (2012). \\

40. J.B.Stephen, L.Bassani, A.Malizia, N.Masetti, and P. Ubertini, Astron. Telegram 11340 (2018). \\

41. D.Stern, R.J.Assef, D.J.Benford, A.Blain, R.Cutri, A. Dey, P. Eisenhardt, R. L. Griffith, et al., Astrophys. J. {\bf 753}, 30 (2012). \\

42. J.Tomsick, S.Chaty, J.Rodriguez, etal., Astrophys. J. {\bf 701}, 811 (2009). \\

43. J. A. Tomsick, R. Krivonos, Q. Wang, A. Bodaghee, S. Chaty, F. Rahoui, J. Rodriguez, and F. M. Fornasini, Astrophys. J. {\bf 816}, 14 (2016). \\

44. W. Wegner, Mon. Not. R. Astron. Soc. {\bf 374}, 1549 (2007). \\
45. W.Wegner, Acta Astron. {\bf 64}, 261 (2014). \\

46. C. Winkler, T.J.-L. Courvoisier, G. di Cocco, N. Gehrels, A. Gimenez, S. Grebenev, W. Hermsen, J. M. Mas-Hesse, et al., Astron. Astrophys. {\bf 411}, L1 (2003). \\

47. Y. Yan, A. Esamdin, and Y. Hong Xing, Sci. China Phys., Mech. Astron. {\bf 53}, 1726 (2010).\\

\bigskip
{\it Translated by V. Astakhov}

\end{document}